\documentclass[a4paper]{jpconf}
\usepackage{graphicx}
\begin{document}
\title{Magnetic-field effects on the charge-spin stripe order in La-214 high-$T_{\rm c}$ cuprates}

\author{T. Adachi$^1$, K. Omori$^1$, T. Kawamata$^1$, K. Kudo$^2$, T. Sasaki$^2$, \\N. Kobayashi$^2$, Y. Koike$^1$}

\address{$^1$ Department of Applied Physics, Tohoku University, 6-6-05 Aoba, Aramaki, Aoba-ku, Sendai 980-8579, Japan}
\address{$^2$ Institute for Materials Research, Tohoku University, 2-1-1 Katahira, Aoba-ku, Sendai 980-8577, Japan}

\ead{adachi@teion.apph.tohoku.ac.jp}

\begin{abstract}
Magnetic-field effects on the charge-spin stripe order in La-214 high-$T_{\rm c}$ cuprates have been investigated from measurements of the in-plane electrical-resistivity, $\rho_{\rm ab}$. 
In La$_{2-x}$Ba$_x$CuO$_4$ with $x = 0.10$ and La$_{2-x}$Sr$_x$CuO$_4$ with $x = 0.115$ where the incommensurate charge peaks are weak and unobservable in zero field in elastic neutron-scattering measurements, respectively, the normal-state value of $\rho_{\rm ab}$ at low temperatures markedly increases with increasing field up to 27 T. 
For La$_{2-x}$Ba$_x$CuO$_4$ with $x = 0.11$ and Zn-substituted La$_{2-x}$Sr$_x$Cu$_{1-y}$Zn$_y$O$_4$ with $x = 0.115$ and $y = 0.02$ where the charge stripe order is fairly stabilized in zero field, on the other hand, the increase in $\rho_{\rm ab}$ with increasing field is negligibly small. 
In conclusion, when the charge-spin stripe order is not fully stable in zero field, magnetic field operates to stabilize the charge-spin stripe order. 
The value of $\rho_{\rm ab}$ increases with increasing field depending on the stability of the charge stripe order. 
\end{abstract}

\section{Introduction}
These years, magnetic-field effects on the so-called charge-spin stripe order [1] in the La-214 high-$T_{\rm c}$ cuprates have attracted much attention.
From elastic neutron-scattering measurements in La$_{2-x}$Sr$_x$CuO$_4$ (LSCO) with $x = 0.10$, it has been found that the intensity of the incommensurate magnetic peaks corresponding to the formation of the spin stripe order is enhanced with increasing field parallel to the c axis [2], suggesting the stabilization of the spin stripe order by the application of magnetic field.
For LSCO with $x = 0.12$, on the other hand, the enhancement of the incommensurate magnetic peaks by the application of magnetic field is small [3], which is due to the fair stabilization of the spin stripe order even in zero field around $x\sim1/8$.
As for magnetic-field effects on the charge stripe order, elastic neutron-scattering measurements for La$_{1.6-x}$Nd$_{0.4}$Sr$_x$CuO$_4$ (LNSCO) with $x = 0.15$, in which the charge-spin stripe order is stabilized in the tetragonal low-temperature (TLT) structure (space group: $P4_2/ncm$), have revealed no enhancement of the incommensurate charge peaks [4]. 

In order to investigate magnetic-field effects on the charge-spin stripe order from the viewpoint of transport properties, we previously measured the in-plane electrical resistivity, $\rho_{\rm ab}$, in magnetic fields parallel to the c axis for La$_{2-x}$Ba$_x$CuO$_4$ (LBCO) with $x = 0.08, 0.10, 0.11$ and LNSCO with $x = 0.12$ [5,6]. 
It has been found for LBCO with $x = 0.10$ in the TLT phase that normal-state values of $\rho_{\rm ab}$ at low temperatures markedly increase by the application of magnetic field. 
Since the incommensurate charge peaks are weaker at $x=0.10$ than at $x\sim1/8$ in LBCO [7,8], the result suggests that the charge stripe order is stabilized by the application of magnetic field, resulting in the change to an insulating behavior of $\rho_{\rm ab}$. 

In this paper, we investigate magnetic-field effects on $\rho_{\rm ab}$ in other La-214 systems, namely, La$_{2-x}$Sr$_x$Cu$_{1-y}$Zn$_y$O$_4$ (LSCZO) with $x = 0.115$ and $y = 0 - 0.02$. 
We discuss effects of nonmagnetic impurities and magnetic field on the charge-spin stripe order, using our previous results of LBCO with $x = 0.08, 0.10, 0.11$ also [5,6]. 

\section{Experimental}
Single crystals of LSCZO with $x = 0.115$ and $y = 0$, 0.01, 0.02 were grown by the traveling-solvent floating-zone method under flowing oxygen gas of 4 bar. 
The Sr and Zn contents were estimated by the inductively-coupled-plasma (ICP) measurements. 
Measurements of $\rho_{\rm ab}$ were performed in magnetic fields up to 27 T by a standard dc four-probe method, using a superconducting (SC) magnet below 15 T and a hybrid magnet above 15 T at the High Field Laboratory for Superconducting Materials (HFLSM), IMR, Tohoku University. 

\section{Results and Discussion}
\begin{figure}
\begin{center}
\includegraphics[width=0.7\linewidth]{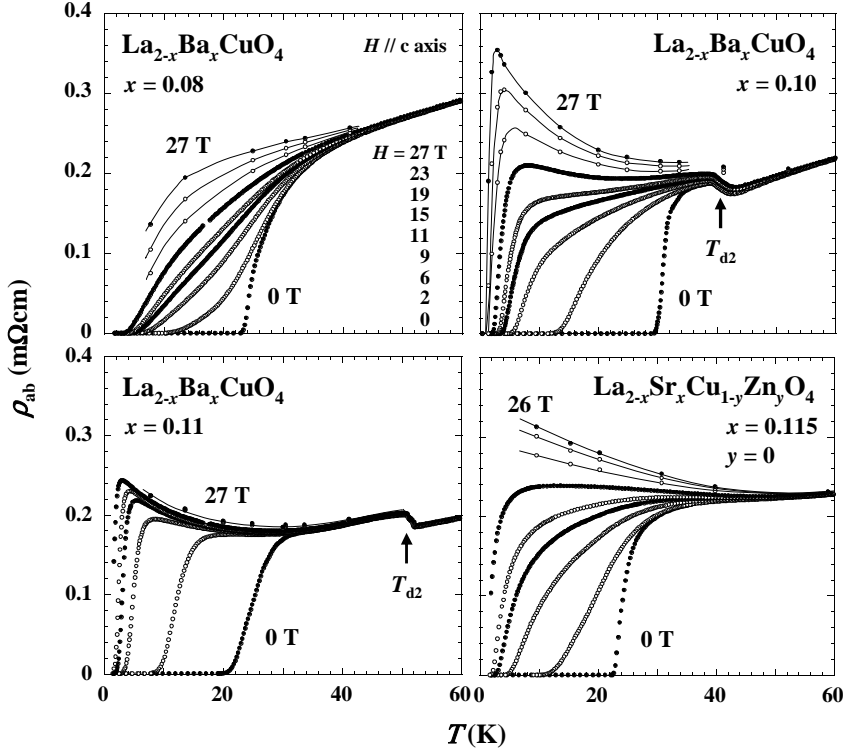}
\end{center}
\caption{\label{label}Temperature dependence of the in-plane electrical resistivity, $\rho_{\rm ab}$, in various magnetic fields parallel to the c axis for La$_{2-x}$Sr$_x$Cu$_{1-y}$Zn$_y$O$_4$ with $x = 0.115$ and $y=0$. The data of La$_{2-x}$Ba$_x$CuO$_4$ with $x = 0.08, 0.10, 0.11$ are also plotted for reference [5,6]. The temperature at which $\rho_{\rm ab}$ jumps corresponds to the structural phase transition temperature between OMT (see the text) and TLT phases, $T_{\rm d2}$.}
\end{figure}

Figure 1 shows the temperature dependence of $\rho_{\rm ab}$ in various magnetic fields for LSCZO with $x = 0.115$ and $y=0$ and also shows our previous results of LBCO with $x = 0.08, 0.10, 0.11$ [5,6].
For LBCO with $x=0.08$, the so-called fan-shape broadening is observed up to 27 T, which is characteristic of the underdoped high-$T_{\rm c}$ cuprates with large SC fluctuations. 
For LBCO with $x=0.10$, the SC transition shows broadening in low fields and it shifts to the lower temperature side in parallel in high fields. 
Moreover, the normal-state value of $\rho_{\rm ab}$ at low temperatures increases with increasing field up to 27 T and an insulating behavior is observed in high fields.
These suggest that the charge stripe order tends to be stabilized by the application of magnetic field up to 27 T. 
For LBCO with $x=0.11$, on the other hand, the SC transition shows a parallel shift with increasing field and the increase in $\rho_{\rm ab}$ with increasing field is negligibly small. 
Similar magnetic-field effects on $\rho_{\rm ab}$ have been observed for LNSCO with $x = 0.12$ [5].
The robustness of $\rho_{\rm ab}$ against the applied magnetic field suggests that the stripe order is fairly stabilized even in zero field around $x = 1/8$.

As for LSCZO with $x = 0.115$ and $y=0$, the broadening behavior of the SC transition at 2 T changes to be a parallel-shift behavior at 9 T and the normal-state value of $\rho_{\rm ab}$ at low temperatures increases with increasing field up to 26 T, resulting in an insulating behavior in high fields.
Moreover, it is found that the overall behavior of the temperature dependence of $\rho_{\rm ab}$ in LSCO with $x=0.115$ and $y=0$ in each magnetic field is quite similar to that observed in LBCO with $x=0.10$.
In elastic neutron-scattering measurements in zero field for LSCO with $x\sim1/8$, no incommensurate charge peaks have been observed in the orthorhombic mid-temperature (OMT) structure (space group: {\it Bmab}) [9]. 
Therefore, it is possible that the charge stripe order is stabilized by the application of magnetic field even for LSCZO with $x=0.115$ and $y=0$. 

\begin{figure}
\begin{center}
\includegraphics[width=0.9\linewidth]{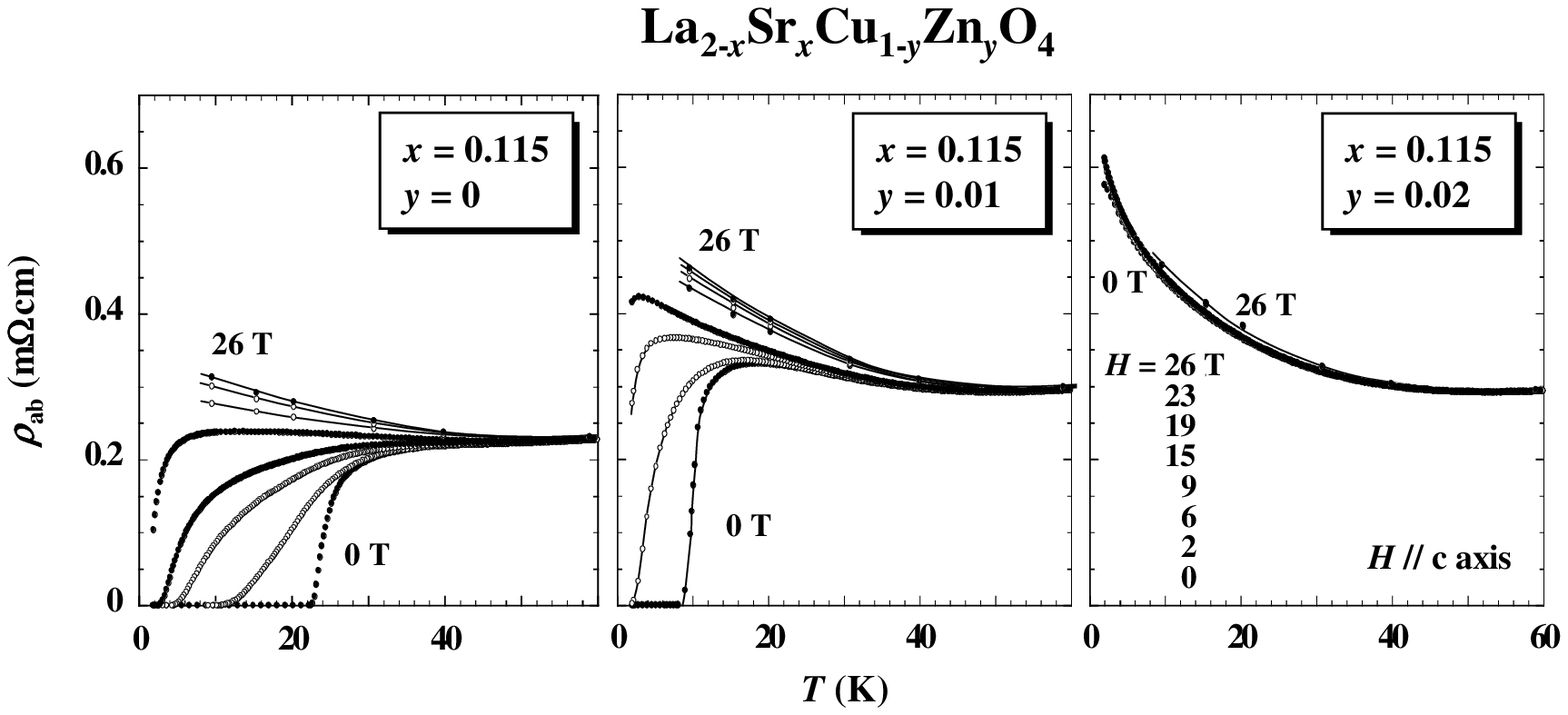}
\end{center}
\caption{\label{label}Temperature dependence of the in-plane electrical resistivity, $\rho_{\rm ab}$, in various magnetic fields parallel to the c axis for La$_{2-x}$Sr$_x$Cu$_{1-y}$Zn$_y$O$_4$ with $x=0.115$ and $y=0$, 0.01, 0.02.}
\end{figure}

Figure 2 displays the temperature dependence of $\rho_{\rm ab}$ in various magnetic fields for LSCZO with $x=0.115$ and $y=0$, 0.01, 0.02.
For $y=0.01$, $\rho_{\rm ab}$ increases with increasing field at low temperatures and tends to be saturated at $\sim$ 26 T.
For $y=0.02$, on the other hand, the increase in $\rho_{\rm ab}$ by the application of magnetic field is negligibly small up to 26 T, which is similar to the results of LBCO with $x=0.11$ and LNSCO with $x=0.12$.
These suggest that the enhancement of $\rho_{\rm ab}$ with increasing field at low temperatures becomes weak with increasing Zn concentration and almost disappears for $y=0.02$, which is consistent with our previous results of the thermal-conductivity measurements in magnetic fields for LSCZO around $x=0.115$ [10,11]. 

It is well known that the normal-state value of $\rho_{\rm ab}$ in zero field increases with increasing Zn concentration and a localized behavior appears at low temperatures in the underdoped high-$T_{\rm c}$ cuprates, as seen in Fig. 2. 
Although the magnetic field also affects carriers to be localized and $\rho_{\rm ab}$ shows an insulating behavior in the underdoped high-$T_{\rm c}$ cuprates [12], one cannot easily understand such significant magnetic-field effects on $\rho_{\rm ab}$ depending on the Zn concentration as shown in Fig. 2. 
Formerly, it has been suggested from zero-field muon-spin-relaxation (ZF-$\mu$SR) measurements for LSCZO with $x=0.115$ that Zn operates to pin the dynamical stripe correlations, leading to the formation of the static stripe order [13]. 
Moreover, the volume fraction of the spin-stripe-ordered region in a sample has been estimated to be $\sim 90$ \% for $y=0$ and 100 \% for $y=0.02$.
Based upon these and the neutron-scattering results [9], the magnetic-field effects on $\rho_{\rm ab}$ for LSCZO with $x=0.115$ can be interpreted as follows.
That is, for $y=0$, the charge stripe order is unstable in zero field, though the spin stripe order is stabilized in a sample.
Accordingly, the charge stripe order is stabilized by the application of magnetic field, resulting in the increase in $\rho_{\rm ab}$ with increasing field.
For $y=0.01$, on the other hand, the charge-spin stripe order is pinned and rather stabilized by Zn, resulting in the weak magnetic-field effect on the stabilization of the stripe order. 
For $y=0.02$, owing to the almost perfect stabilization of the stripe order by Zn, the magnetic-field effect on $\rho_{\rm ab}$ almost disappears. 
Therefore, it is concluded that both the magnetic field parallel to the c axis and Zn operate to stabilize the charge stripe order in La-214 high-$T_{\rm c}$ cuprates.

\section{Summary}
From $\rho_{\rm ab}$ measurements in magnetic fields for LSCZO with $x=0.115$, it has turned out that the normal-state value of $\rho_{\rm ab}$ at low temperatures increases with increasing field for $y=0$ in which the charge stripe order is unstable.
Moreover, it has turned out that the enhancement of $\rho_{\rm ab}$ is negligibly small for $y=0.02$ in which the charge stripe order is fairly stabilized by the substituted Zn.
Taking into account our previous results of $\rho_{\rm ab}$ in magnetic fields for LBCO with $x=0.08$, 0.10, 0.11, these results suggest that (i) the magnitude of the enhancement of $\rho_{\rm ab}$ by the application of magnetic field is in intimate relation with the stability of the charge stripe order in zero field and (ii) both the magnetic field parallel to the c axis and Zn operate to stabilize the charge stripe order in La-214 high-$T_{\rm c}$ cuprates.

\ack{We are indebted to Prof. K. Takada and M. Ishikuro for their help in the ICP analysis.}

\end{document}